\documentclass[12pt,a4paper,twoside]{article}

\usepackage[dvips]{graphics}
\usepackage{cite}

\newcommand{\ycut}   {\ensuremath{y_{\mathrm{cut}}}}
\newcommand{\ybcut}  {\ensuremath{{\bf y_{\mathrm{{\bf cut}}}}}}
\newcommand{\njet}   {\ensuremath{N_{\mathrm{jet}}}}
\newcommand{\epem}   {\ensuremath{\mathrm{e^+e^-}}}
\newcommand{\yij}    {\ensuremath{y_{ij}}}
\newcommand{\ymn}    {\ensuremath{y^{\mathrm{m\leftarrow n}}}}

\newcommand{\vij}    {\ensuremath{v_{ij}}}
\newcommand{\yini}   {\ensuremath{y_{\mathrm{init}}}}
\newcommand{\ymax}   {\ensuremath{y^{\mathrm{max}}_{ij}}}
\newcommand{\ytt}    {\ensuremath{y^{2\leftarrow 3}}}
\newcommand{\ytf}    {\ensuremath{y^{3\leftarrow 4}}}
\newcommand{\nobj} {\ensuremath{N_{\mathrm{obj}}}}
\newcommand{\eeqq}   {\ensuremath{e^+e^-\rightarrow q\bar{q}}}

\newcounter{hours}
\newcounter{minutes}

\newcommand{\Printtime}{%
  \setcounter{hours}{\time/60}%
  \setcounter{minutes}{\time-\value{hours}*60}%
  \ifthenelse{\value{hours}<10}{0}{}\thehours:%
  \ifthenelse{\value{minutes}<10}{0}{}\theminutes}

\begin{document}


%
\begin{titlepage}

%
\begin{center}
{ \large EUROPEAN LABORATORY FOR PARTICLE PHYSICS }
\end{center}
\bigskip

%
\begin{flushright}
  CERN-EP/98-043 \\
  March 11th, 1998
\end{flushright}
\bigskip\bigskip

\begin{center}
{\Huge\bf
The Cambridge jet algorithm:
\\[0.2cm]
features and applications 
}
\end{center}
\bigskip\bigskip
\begin{center}
{\large Stan Bentvelsen\footnote{CERN, European Organisation for
 Particle Physics, CH-1211 Geneva 23, Switzerland} and Irmtraud
 Meyer\footnote{III. Physikalisches Institut, RWTH Aachen, 52056 Aachen,
 Germany}}
\end{center}
\bigskip

%
\begin{abstract}
Jet clustering algorithms are widely used to analyse hadronic events
in high energy collisions. Recently a new clustering method, known as
`Cambridge', has been introduced. In this article we present an algorithm
to determine the transition values of \ycut\ for this clustering
scheme, which allows to resolve any event to a 
definite number of jets in the final state.  We discuss some
particularities of the Cambridge clustering method and compare its
performance to the Durham clustering scheme for Monte Carlo generated
\epem\ annihilation events.
\end{abstract}
\bigskip\bigskip\bigskip
 
%
\begin{center}
{\large Submitted to Eur. Phys. J. C}
\end{center}
\bigskip
 
%
\vspace*{1.5 cm}
%
\vspace*{1 cm}
\begin{center}

\end{center}

\end{titlepage}

%
%
\newpage
\section{ Introduction }
In collider physics, clustering of the experimentally accessible
hadronic final states is used to determine the underlying parton
structure of events.  In \epem\ annihilation the widely known
JADE~\cite{b:jade} and Durham~\cite{b:durham} jet algorithms have
become indispensable in this process, permitting a wide range of important
tests of QCD, allowing refined measurements of electro-weak physics with
hadronic final states and being used in searches for new physics.

Recently a new jet clustering scheme, known as Cambridge, has been
introduced~\cite{b:cambridge}. This scheme is a modification of the
original Durham $k_{T}$-clustering scheme. 
The Cambridge algorithm is designed to minimise the formation of
spurious `junk-jets', jets formed from a multitude of low transverse
momentum particles, unrelated to the underlying parton structure.


For all the above mentioned algorithms, clustering of the final
state is performed iteratively and is terminated at a clustering
specific resolution scale, generically denoted by the resolution
parameter \ycut. By changing the value of \ycut, the final state is
resolved into a varying number of jets.  The Cambridge algorithm involves
three basic components in this iterative process.  It uses an
ordering variable, \vij, a test variable, \yij, and a recombination
procedure. 
In JADE-type jet clustering algorithms
only two basic components are involved, since the ordering variable,
\vij, and the test variable, \yij, are identical.

In this note we review the Cambridge finder and discuss some of its
experimental peculiarities.  
In terms of computing this algorithm is more complex compared to the
JADE and Durham algorithms. Due to the distinction between test and
ordering variables, the sequence of clustering now depends on the
value of \ycut.  We show that the jet multiplicity obtained with this
algorithm is not monotonically decreasing for increasing \ycut, and
that for some events it is impossible to resolve a certain jet
multiplicity.  Therefore the concept of the `transition values in
\ycut' has to be defined more precisely. The transition value at which
the event classification changes from $n$-jets to $m$-jets, when going
to larger values for \ycut, will subsequently be referred to as \ymn\
value.

Next we developed a fast algorithm to obtain the transition
values \ymn\ for the Cambridge finder. Using this algorithm, we
compare results for Monte Carlo generated $\epem \rightarrow q\bar{q}$
events between the Durham and Cambridge finder. We compare their 
performance in determining the size of the hadronization corrections.
As another example, we determine the performance for hadronic decays
of $W^+W^-$ production at LEP2~\cite{b:lep2ww}. Finally we give our
conclusions and cite an address to download our FORTRAN code.

\section{ The Cambridge algorithm }
In JADE-type jet clustering algorithms one iteratively combines
particles to form final state jets. First one introduces a `test
variable' \yij. The pair of two objects $i$ and $j$ with smallest
value for \yij\ is selected and its objects are combined or the
iteration is terminated when $\yij > \ycut$ for all pairs of objects.
For the JADE and Durham algorithms, the test
variables $\yij^J$ and $\yij^D$ are defined respectively as
\begin{eqnarray}
\yij^J & =  & \frac{2 E_i E_j }{E_{vis}^2}(1-\cos\theta_{ij}) \\
\yij^D & =  & \frac{2 \min(E_i^2,E_j^2)}{E_{vis}^2}(1-\cos\theta_{ij})
\end{eqnarray}
where $E_i$ and $E_j$ denote the energies of particles $i$ and $j$ and
$\theta_{ij}$ their opening angle.  Note that we normalise the values
of $\yij^J$ and $\yij^D$ to the visible energy, $E_{vis}$, which is the
sum of energies for all particles observed in the final state.  The
second ingredient is the recombination procedure.  Normally the
$E$-scheme is taken, for which the four-momentum of the resulting
object is simply the sum of the four-momenta of the two objects $p_i$
and $p_j$. 


In contrast to this the Cambridge algorithm involves three basic
components to form the final state jets. The algorithm starts from a
table of \nobj\ primary objects, which is the set of the particles'
four-momenta. It starts clustering the pair of particles with the
smallest opening angle, using the ordering variable \vij.  The
test variable $\yij^D$, which is identical to the one for the Durham
algorithm, decides when the iterative procedure is stopped. It is
subsequently denoted by \yij.  The algorithm proceeds as follows:
\begin{enumerate}
\item If only one object remains, store this as a jet and stop.
\item Select the pair of objects $i$ and $j$ that have the minimal
  value for their ordering variable, \vij, with 
  $\vij=2(1-\cos \theta_{ij})$.
\item Inspect the test variable \yij.
  \begin{itemize}
  \item If $\yij <\ycut$ then combine $i$ and $j$ in a new object
    using the $E$-scheme. Remove particles $i$ and $j$ from the
    table of objects that remain to be combined and add the new
    object with four-momentum $p_i+p_j$.
  \item If $\yij \geq \ycut$ then store the object $i$ or $j$  with the smaller
    energy as a separated jet and remove it from the table. The higher
    energetic object remains in the table.
  \end{itemize}
\end{enumerate}
Removing the softer of two resolved objects, as described in the
last step, is called {\em soft freezing}. It prevents the softer jet
from attracting any extra particles, thereby reducing non-intuitive clustering
effects.

   
\section{ The Cambridge algorithm: an example }
In order to test the various clustering algorithms, we generate Monte
Carlo \eeqq\ events at $\sqrt{s} = 91.2$ GeV with the PYTHIA event
generator~\cite{b:pythia}. The generation includes parton showering
(`parton-level'), and subsequent fragmentation and decays of the final
state (`hadron-level'). The parameters of the Monte Carlo event
generator are adjusted in order to provide an optimal description of
large samples of hadronic $Z^0$ decay data~\cite{b:pythiaopal}.

                                %
                                %
\begin{figure}[!tb]
  \begin{center}
    \resizebox{\textwidth}{!}
    {\includegraphics{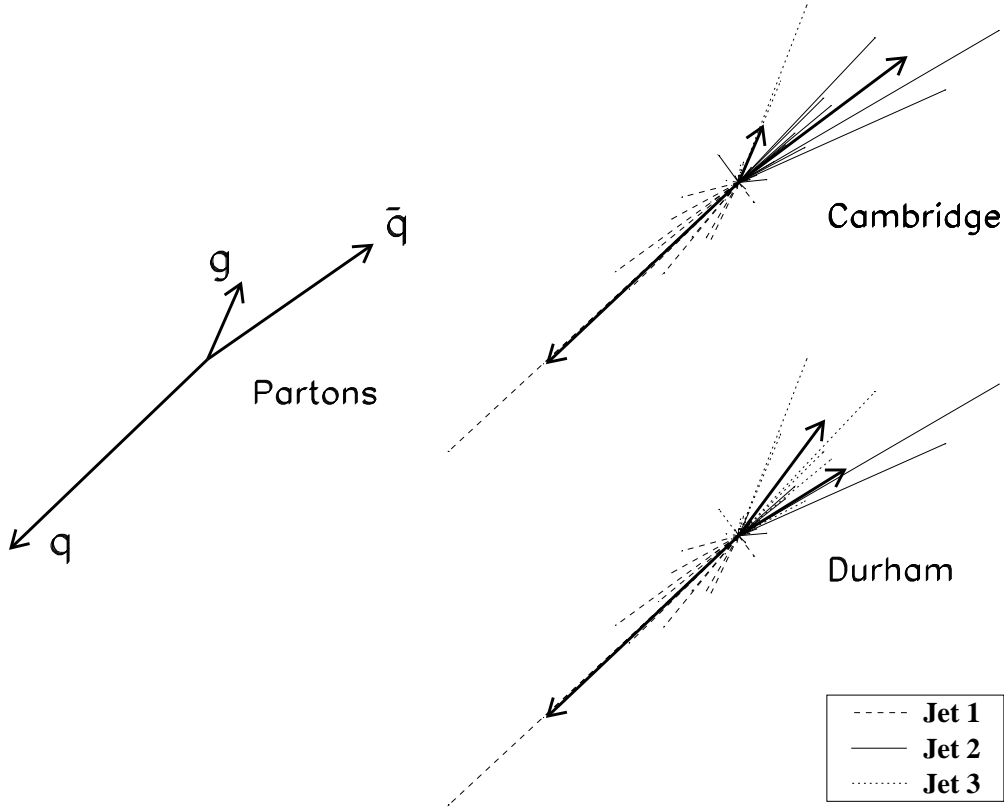}}
    \caption[]
    { {\sl Display, projected onto the $xy$-plane, of a $q\bar{q}$
      generated event at $\sqrt{s} = 91.2$ GeV. 
      The lengths of the
      arrows are proportional to the energies of the objects. The
      jet-axes are indicated by the arrows, left at the parton
      level, right at the hadron level. The particle and jet three-momenta
      are shown for the Cambridge and Durham algorithms
      separately. Particle association to jets is indicated with various
      line-styles. The length of the parton and jet axes are scaled down
      by a factor of four. }}
    \label{f:fourv}
  \end{center}
\end{figure}
To illustrate the differences between the Cambridge and Durham finders
we present in Figure~\ref{f:fourv} the three-momenta of a typical event
projected onto the $xy$-plane. The underlying parton level is shown in
the figure by the thick arrows and consists of a quark $q$ recoiling against 
a $\bar{q}g$ system,
with the gluon being relatively soft. 

At the hadron level, the event is clustered again to three final state jets,
both with the Durham and Cambridge algorithm. The final jets are
indicated by thick arrows, and the association of particles to the three
jets is indicated by various line styles.

In this example one clearly observes the positive effect of soft freezing on
the hadronization corrections.  In the Cambridge algorithm the soft
gluon jet is separated and classified as a final state jet. Most
particles in the hemisphere are assigned in an intuitive way to the quark
jet. The three final jets closely resemble the underlying parton
structure. In contrast to this, in the Durham algorithm
more particles are clustered around the
soft gluon, so that the gluon jet becomes even more energetic than the
quark jet.  It is obvious that in this example the final state found for
the Cambridge algorithm resembles the parton structure more than the
Durham algorithm. 

                                %
                                %
\begin{figure}[!tb]
  \begin{center}
    \includegraphics{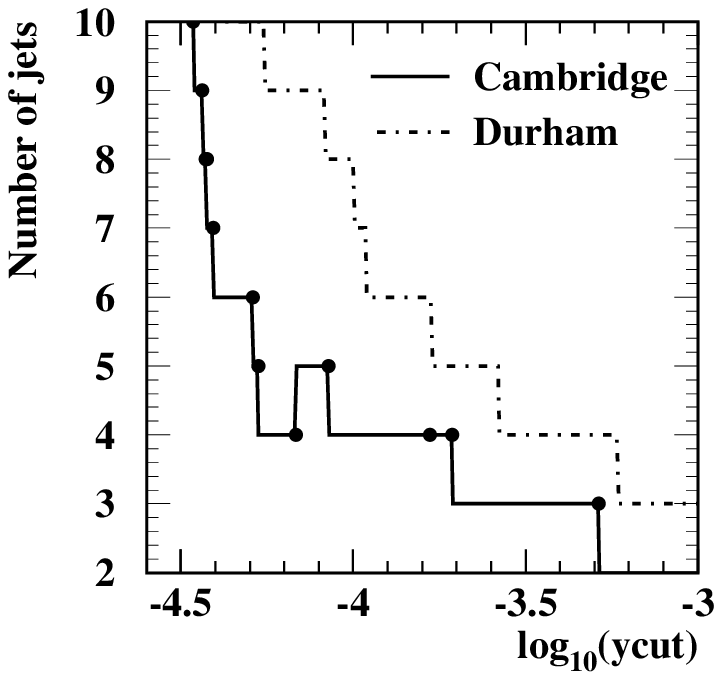}
    \includegraphics{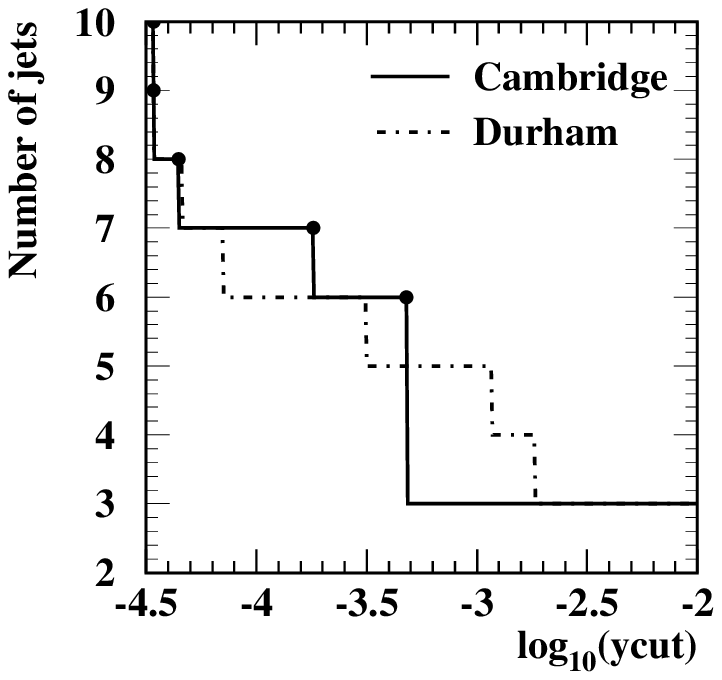}
    \caption[]
    { {\sl Example of the jet multiplicity, \njet\, as function of \ycut\
      for two Monte Carlo generated events. The dash-dotted line is for the
      Durham algorithm, the full line for the Cambridge algorithm.
      The points mark the \ymn\ transition values. 
      }}
    \label{f:jetm}
  \end{center}
\end{figure}
As an illustration of some of the peculiarities of the Cambridge
algorithm, in Figure~\ref{f:jetm} we present the jet multiplicity as
function of \ycut\ for two events. In the figures the dashed lines
correspond to the Durham algorithm, whereas the full lines correspond
to the Cambridge clustering.  For the Durham algorithm, the jet
multiplicity is decreasing monotonically for increasing \ycut. In
addition, each event can be resolved into each jet-multiplicity \njet,
with $1 \leq \njet \leq \nobj$.

In the Cambridge finder the situation is a little more complex, as can be 
seen in the same figure. In the left plot of Figure~\ref{f:jetm} an
example is given where
\begin{itemize}
\item the jet multiplicity is not monotonically decreasing for
increasing \ycut.
\end{itemize}
In this example, at the resolution $\ycut \sim 10^{-4.3}$, the
jet-multiplicity decreases from 6 to 5 to 4 when \ycut\ increases, but
then increases from 4 to 5 again. At $\ycut \sim 10^{-4.1}$ the
multiplicity decreases from 5 to 4.  At $\ycut \sim 10^{-3.8}$ a
situation occurs where the jet multiplicity does not change from being
4, but the four final state jets change their four-momenta.  The jet
configuration of the two 5-jet states and three 4-jet state are all
different.

In the right plot of Figure~\ref{f:jetm} we show an example where
\begin{itemize}
\item it may not be possible to resolve the event into a certain $n$-jet
final state.
\end{itemize}
In this example, at $\ycut \sim 10^{-3.3}$, the event changes from being
classified as a 6 jet event to a 3 jet event. For this event, it
is impossible to choose a value for \ycut\ such that the event is
resolved into a 4 or a 5 jet configuration.

As a last `peculiarity' of the Cambridge jet clustering we consider
the particle to jet association. For the JADE and Durham algorithms,
when crossing a transition value in \ycut\ towards higher values, two
of the jets are merged into one new jet while all other jets are left
untouched.  The resulting new jet consists of exactly all particles
that belonged to the two merged jets, and the subjet history of jets
can be traced unambiguously.  In the Cambridge algorithm this need not
be the case, since the sequence of recombination may be different for
different values of \ycut.  It thus can happen that the
particle contents of a jet at a given value for \ycut\ does not match
the sum of the particle contents of two resolved jets at lower values
of \ycut.

For many applications it is essential to obtain the transition values \ymn.
For example, in previous studies of \epem\ annihilation
data the value of \ytt\ was analysed in order to obtain
$\alpha_s(Q^2)$~\cite{b:alphas}.  In other studies all events were
classified as four~\cite{b:tripleglue} or five~\cite{b:fivej} 
jets and their angular correlations were studied in
order to probe the non-Abelian nature of QCD. In
current studies at \epem\ annihilation energies reached by the LEP2
programme~\cite{b:lep2ww}, events also have to be clustered to four jets
in order to determine the $W$-boson characteristics in the hadronic
decays of $W^+W^-$ pairs.  Therefore, an algorithm
to obtain all transition values \ymn\ with full information on the
particle to jet association is highly desirable.


\section{ Transition values of \ybcut }

In the JADE and Durham algorithm, the sequence of clustering of an
event can be determined once and completely, and is independent of the
value of \ycut.  From this clustering information about jet
multiplicities, four-momenta and jet-particle association, can
subsequently be retrieved for any value of \ycut. This is the strategy
used in the {\tt KTCLUS}\cite{b:ktclus} and {\tt YKERN}~\cite{b:ykern}
packages. The final jet configuration is identical for all values of \ycut\
between two subsequent transition values. At the transition value
$y^{n\leftarrow n+1}$, the event flips from a $n+1$-jet to a $n$-jet
configuration.  The transition values for the JADE and Durham
algorithms are ordered in \ycut. Using the transition values one can
select a value for \ycut\ such that the event is resolved into the
required number of jets.

In contrast to this, the clustering sequence in the Cambridge
algorithm depends on the value of \ycut\ because it
distinguishes between ordering and testing variables.  It is therefore
no longer straightforward to calculate the transition values. In
general, at the transition values $y^{m\leftarrow n}$ the event can
flip between a $n$-jet configuration to a $m$-jet configuration where
$n$ and $m$ are not necessarily consecutive.  
As it is important to obtain the values for \ymn, 
it was suggested in~\cite{b:cambridge}
to perform a binary search in \ycut\ to determine these
transition values, by repeated evaluation of the clustering.  This
proposal is not completely satisfactory since such a search has an
intrinsic limited precision, might skip over several transition values
 and becomes very computing time intensive.

We have developed a method to determine the transition values of \ycut\ for the
Cambridge finder exactly, as follows. While performing the clustering
at a particular value of \ycut, denoted by \yini, we keep track of the
maximum value of \yij, between any two objects $i$ and $j$ encountered in
this process, with \yij\ being always smaller than
\yini. By construction this maximum value, which we denote by
\ymax, is smaller than \yini.  We now note that for any value of
$\ycut \in \left[ \ymax, \yini \right) $, 
the Cambridge algorithm will follow the
same clustering sequence.  Only when the cluster algorithm is
performed with a value \ycut\ smaller than \ymax, the
condition $\yij \geq \ycut$ is satisfied at least once more and the
subsequent clustering sequence may change completely.  The value
\ymax\ is therefore one of the \ycut\ transition values. Note that
the clustering may also change completely for values of \ycut\ larger than \yini.

These observations can be utilised to scan the complete region of \ycut.
We therefore start by clustering the complete event to a one-jet
configuration by chosing $\ycut = 1$ in the first step. After this
step one iteratively repeats the clustering to calculate smaller and
smaller values of \ymax\ at which the clustering changes, and thereby
calculates smaller and smaller transition values. The process terminates
if either the number of resolved jets equals the number of
input objects or if the desired number of jets is resolved. 
To summarise:
\begin{enumerate}
\item Start with value $\yini = 1$ and set $\ycut = \yini$.
\item Perform the Cambridge jet clustering for \nobj\ objects.
  During the clustering, keep track of all
  values of \yij\ between all objects $i$ and $j$, and determine their
  maximum value, \ymax.
\item Store the value of \ymax, the number of jets, $n$,
  their four-momenta and the jet-particle association.
  The clustering for $\ycut > \ymax$ is now completely determined. 
\item The algorithm stops if:
  \begin{itemize}
    \item The number of resolved jets equals the number of input
      objects, $n= \nobj$. Then the event is classified completely and
      the algorithm necessarily stops.
    \item The desired number of jets or a preset lower limit in \ycut\
      is reached, and the algorithm is stopped.
  \end{itemize}
\item Set $\ycut=\ymax$ and go to step 2.
\end{enumerate}
Once this process has been performed, all information about
the clustering is accessible without any appreciable additional
computing time. The total amount of computing time is proportional to
the desired jet-multiplicity. For example, to study four jet final
states with the Cambridge finder requires approximately four times as
much computing time compared to the Durham or JADE algorithms.
\\[0.6cm] 
Instead of the top-down approach for which the clustering
starts at $\ycut =1$ as explained above, a bottom-up approach is in
principle also possible.  One may implement the bottom-up approach by
starting the clustering at the lowest value $\ycut = 0$.  For \nobj\
given at the start of the clustering, one finds the pair of objects
with smallest value \vij\ (corresponding to the pair closest in angle)
and determines the corresponding value for \yij. Then the two possible
cases are considered: one in which the softer object is frozen, the
other in which the two objects are combined. In both cases the number
of objects that remain to be combined is reduced by one.  This
combinatorical procedure is subsequently continued, and {\em all}
possible clustering sequences are listed.  The procedure terminates
when only one object remains.

From the corresponding values for \yij, saved for each step, one can
deduce the final transition values \ymn\ and the jet configuration
associated to them.  Note that the number of possible clustering
sequences is proportional to $2^{\nobj+1}$, which limits the practical
use of the bottom-up approach.

\section{ Monte Carlo results }
\subsection*{ Jet finder comparison and hadronization corrections }
\begin{figure}[!p]
  \begin{center}
    \resizebox{\textwidth}{!}  {\includegraphics{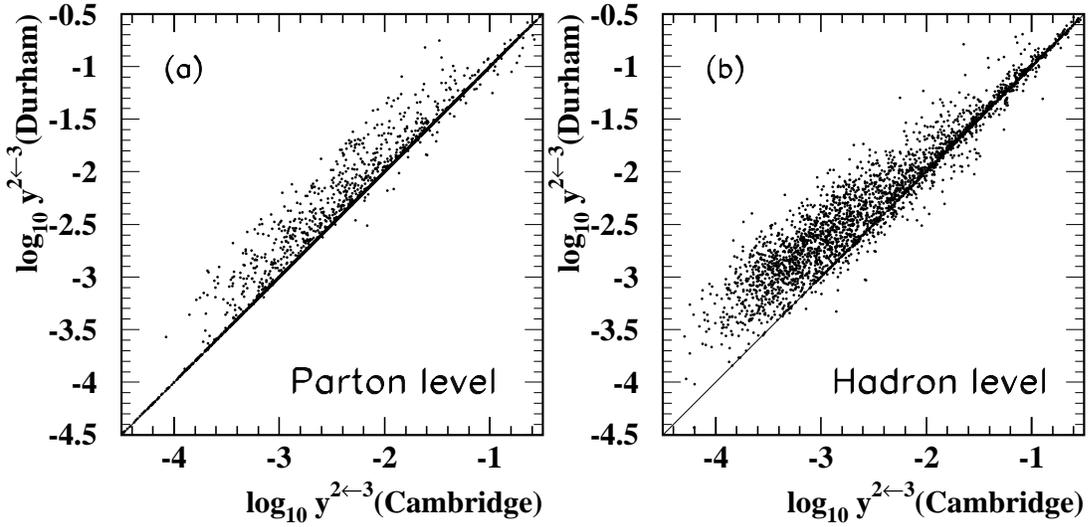}}
    \caption[]
    { {\sl Monte Carlo generated events at $\sqrt{s} = 91.2$ GeV 
       using PYTHIA~\cite{b:pythia}. In a)
      we present the correlation between \ytt\ calculated using the
      Cambridge and \ytt\ calculated using the Durham algorithm, at
      the parton level. Note that the majority of events
      have identical values of \ytt\ for both algorithms. 
      In b) the same is shown for the hadron level.  }}
    \label{f:mctest1}
  \end{center}
\end{figure}

With the transition values \ymn\ defined both for the Cambridge and
Durham algorithms, we compare, as an example, the values for \ytt.
  In~\cite{b:comp} similar studies have been performed to compare
the performance of the Durham and JADE algorithms.  In all the
following we will define the region with the highest value for \ycut\
as the nominal \ycut\ region.  Here, in Figure~\ref{f:mctest1}a we
show the correlation of the Cambridge and Durham algorithms for \ytt\
at the parton level, at the end of the PYTHIA parton shower. For most
of our generated events, the obtained values for \ytt\ are identical
for the Cambridge and the Durham algorithms (approximately 75\% of the
events are found on the line in the figure). For a small fraction of
events, the value obtained with the Cambridge algorithm is smaller
compared to the value for the Durham algorithm. At the hadron level,
as shown in Figure~\ref{f:mctest1}b, the values at low \ytt\ obtained
using the Cambridge algorithm are smaller for almost all events, but
become similar for the two algorithms for increasing values of \ytt.
At the hadron level, approximately 15\% of the events have identical values 
for \ytt\ for both algorithms.

\begin{figure}[!p]
  \begin{center}
    \resizebox{\textwidth}{!}  {\includegraphics{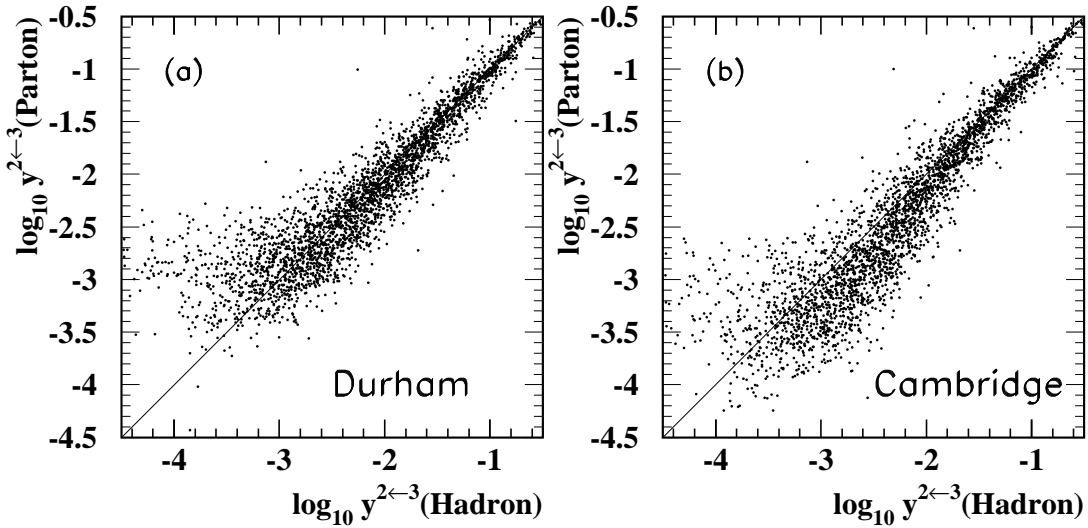}}
    \caption[]
    { {\sl Monte Carlo generated events at $\sqrt{s} = 91.2$ GeV 
      using PYTHIA~\cite{b:pythia}. In a) we
      present the correlation between \ytt\ calculated at the
      parton level and \ytt\ calculated at the hadron level, using the
      Durham algorithm.  In b) the
      same is shown for the Cambridge algorithm. 
      }}
    \label{f:mctest2}
  \end{center}
\end{figure}

Next, in Figure~\ref{f:mctest2}a and \ref{f:mctest2}b, we compare the
hadronization corrections for the Cambridge and Durham algorithms.  We
present the correlation between the transition values \ytt\ calculated
at the hadron and at the parton level, for both.  The line indicates
the ideal case for which equal values for \ytt\ at both levels are
found.  For the Durham algorithm the difference in \ytt\ at the parton
and hadron level is small.  When going to lower \ytt\ values, the
distribution broadens and shifts toward smaller \ytt\ values at the
hadron level.  For the Cambridge algorithm, at high values of \ytt\
the parton and hadron level correlation is similar to the one for the
Durham algorithm. Whereas, when going to smaller values for \ytt, the
values at the hadron level get increasingly larger with respect to the
parton level values.  The width of the distribution is similar to that
for the Durham algorithm. In order to quantify the differences, we
calculated the mean of the logarithmic ratio of the \ytt\ values for
the parton level and the hadron level: this value equals
$0.232\pm0.002$ for the Durham algorithm, and $0.257\pm0.002$ for the
Cambridge algorithm, which indicates that the overall hadronization
corrections for the Durham algorithm are $\sim 10$\% smaller than for
the Cambridge algorithm. Note however that the hadronization
corrections do not only depend on the jet algorithm but also on the
hadronization model used.

%
%

\begin{figure}[!tb]
  \begin{center}
    \resizebox{\textwidth}{!}  {\includegraphics{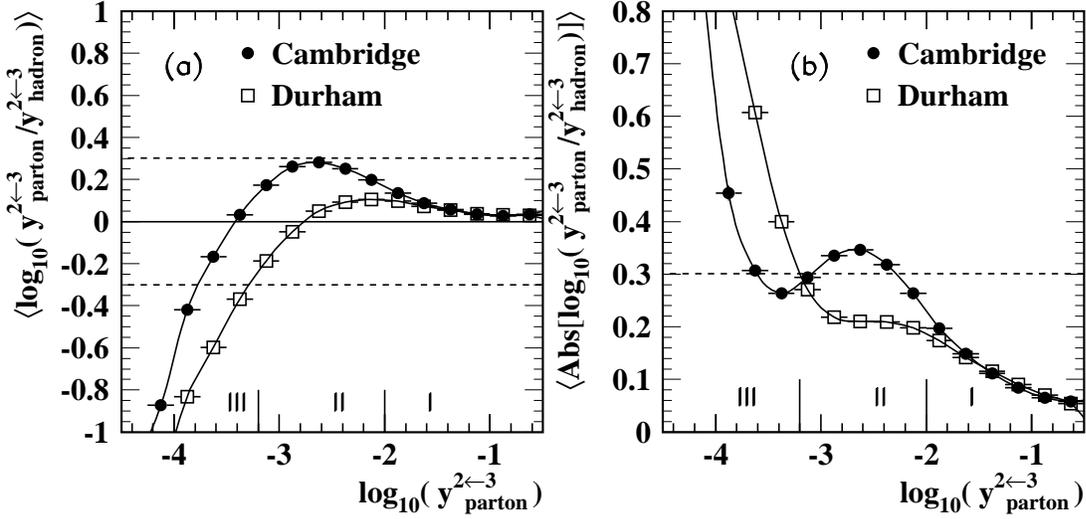}}
    \caption[]
    { {\sl In a) we show for both the Cambridge and the Durham
      algorithms the mean logarithmic ratio between \ytt\ calculated
      at the parton and the hadron level, as function of \ytt\ at the
      parton level, using the PYTHIA event generator. 
      Identical values for \ytt\ are represented by the
      full horizontal line. The dashed lines correspond to a deviation between
      the parton and hadron level by a factor of two.  In b) we
      present the absolute value of the mean as calculated in a).
      The three regions indicated in both figures by $I$, $II$ and $III$
      are discussed in the text.
     }}
    \label{f:mctest3}
  \end{center}
\end{figure}

The mean hadronization corrections can be studied more directly, as a
function of \ycut, from plots as presented in Figures~\ref{f:mctest3}a
and \ref{f:mctest3}b. In Figure~\ref{f:mctest3}a we show the mean of
the logarithmic ratio of the values \ytt\ for the parton and hadron
level, $\left< \log_{10}( \ytt_{parton} / \ytt_{hadron} ) \right>$, as
a function of the transition value at the parton level.  When
calculating the mean deviation between hadron and parton level for
each value \ytt, the contribution to the hadronization corrections for
many events may cancel. To exclude effects due to cancellation we
present in Figure~\ref{f:mctest3}b the size of the hadronization
corrections. We show the mean absolute difference of \ytt\ calculated on the
parton and at the hadron level, $\left< \mbox{Abs} \left( \log_{10}(
\ytt_{parton} / \ytt_{hadron} )\right) \right>$, as a function of \ytt\
calculated at the parton level.

\begin{figure}[!tb]
  \begin{center}
    \resizebox{\textwidth}{!}  {\includegraphics{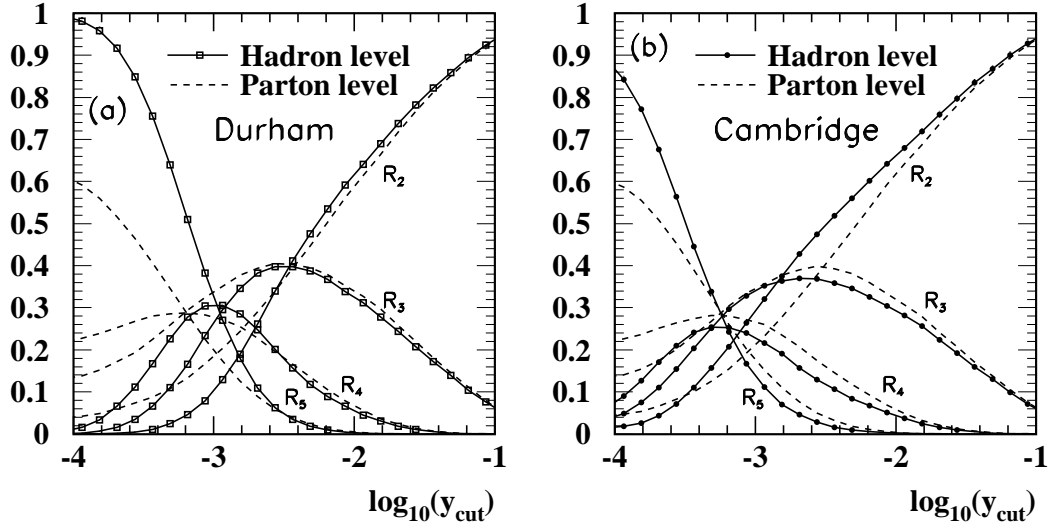}}
    \caption[]
    { {\sl In a)
      we present the relative jet production rates $R_n$, as a function of
      \ycut, for the Durham algorithm, using the PYTHIA event generator.
      The full lines with the
      points correspond to the hadron level and the dashed line
      correspond to the parton level. 
      For $R_5$ we summed all contributions for jet
      multiplicities of 5 and larger. In b) the same is shown for the
      Cambridge algorithm. 
      } }
    \label{f:mctest4}
  \end{center}
\end{figure}


To compare the performance of the two jet finders we distinguish in
Figure~\ref{f:mctest3}a and \ref{f:mctest3}b three regions in \ytt,
denoted by $I$, $II$ and $III$.  In region $I$, for values of \ytt\
above $10^{-2}$, the hadronization corrections are small and
comparable for both algorithms. A fraction of about 37\% of our
generated events belongs to this region.  

Region $II$ is defined for values of \ytt\ between $10^{-3.2}$ and
$10^{-2}$.  In this region differences between the two algorithms
occur.  The mean deviation for the Durham algorithm reaches a maximum
of about 20\%, and vanishes at $\ytt\sim 10^{-2.8}$, as can be seen in
Figure~\ref{f:mctest3}a. However, Figure~\ref{f:mctest3}b shows that
this decrease is due to cancellations and that the absolute
hadronization corrections increase at $\ytt\sim 10^{-2.8}$.  For the
Cambridge algorithm a different behaviour is observed.  The mean
deviation reaches a maximum of 100\% for this algorithm, at $\ytt\sim
10^{-2.6}$, implying large hadronization corrections.  The absolute
hadronization corrections for the Cambridge algorithm, as shown in
Figure~\ref{f:mctest3}b, reach a maximum at about $10^{-2.6}$, and
then decrease until the value for the Durham algorithm is reached. In
the whole region $II$, where about 49\% of our generated events can be
found, hadronization corrections for the Cambridge algorithm are
significantly larger than for the Durham algorithm.

In region $III$, for values of \ytt\ below $10^{-3.2}$, the figures
show that the hadronization corrections are large for both algorithms and
that they increase rapidly towards smaller values of \ytt. 
The corrections for the Cambridge algorithm are smaller compared to
the Durham algorithm. However, only about 14\% of our generated events
can be found in region $III$.

Our analysis of the hadronization corrections for \ytt\ shows that the
Cambridge algorithm performs clearly better only in the region of low
\ytt\ values (region $III$).  This region contains 14\% of the events
and the corrections there are large for both algorithms. In all other
regions the Durham finder performs equally well (region $I$) or even
significantly better (region $II$). These two regions contain a
fraction of 85\% of our generated $\epem\rightarrow q\bar{q}$ events.
These basic tendencies of the hadronization corrections are also found
when studying the transition value \ytf.  Note that for very low
values of \ycut\ the approximations used for the implemention of QCD
in PYTHIA might not give a reliable description of the hadronization
process. Therefore, for very low values of \ycut\ the jets returned by
the Cambridge algorithm may correspond closer to the underlying parton
structure than can be shown in these Monte Carlo studies.  
\\[0.4cm]
Classical tests of QCD rely on relative production rates for multijet
hadronic decays, defined as $R_n = \sigma_n /
\sigma_{tot}$~\cite{b:ykern}.  In Figures~\ref{f:mctest4}a and
\ref{f:mctest4}b we present the relative production rates for two,
three, four, and five or more jet final states, for the hadron level
and the parton level. For these figures we used the same set of
$\epem\rightarrow q\bar{q}$ events as before.  In
Figure~\ref{f:mctest4}a the performance for the Durham algorithm is
shown.  For all \ycut\ values between one and approximately
$10^{-2.8}$ the hadron and parton level agree reasonably well. When
going to smaller values of \ycut\ the curves for the two levels
increasingly deviate.  For the Cambridge algorithm the hadronization
corrections are larger for the region in \ycut\ where the Durham
algorithm performs well, between \ycut\ of one and approximately
$10^{-2.8}$. However, when going to lower values of \ycut\, the
differences between hadron and parton level are, for most jet
multiplicities, smaller than for the Durham algorithm.

To summarise our investigations of the hadronization corrections, we
conclude that the Durham algorithm provides smaller hadronization
corrections for a large region in \ycut. 

In~\cite{b:cambridge}, the hadronization corrections for the mean jet
multiplicity, $\left< n_{jet} \right>=\sum_{1}^{\infty} n R_n$, was
studied.  There it was found that hadronization corrections for the
Cambridge and Durham algorithms are small for values of $\ycut >
10^{-3.2}$.  Our Figures~\ref{f:mctest4}b show that for these values
of \ycut\ the hadronization corrections for each jet production rate,
$R_n$, are sizable for the Cambridge algorithm, whereas for the Durham
algorithm they are small. The small hadronization corrections found
for the Cambridge algorithm in the study of the mean jet rate $\left<
n_{jet} \right>$ are due to fortuitous cancellations in the individual jet
production rates.

\subsection*{ Multiple and impossible jet multiplicities }
As already indicated, the transition values \ymn\ in the
Cambridge algorithm need not be the transition between two consecutive
jet-multiplicities.  Several intervals in \ycut\ may lead to
the same jet-multiplicity, and they have in general different jet
four-momenta. Secondly, it need not always be possible to cluster the
event to any required jet multiplicity.

                                %
                                %
\begin{figure}[!tb]
  \begin{center}
    \resizebox{\textwidth}{!}  {\includegraphics{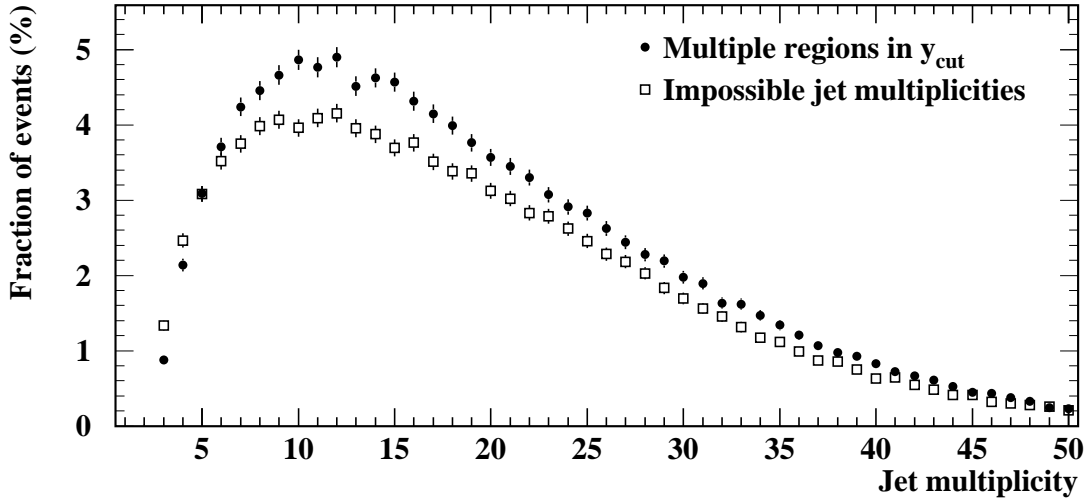}}
    \caption[]
    { {\sl The full points show, as a function of $n$-jet, the
      fraction of events for which multiple $n$-jet final states could
      be reconstructed. This is done for a set of $\epem \rightarrow
      q\bar{q}$ generated events.  The open points show, as a function
      of $n$-jet, the fraction of events for which no $n$-jet final
      state could be reconstructed.  All points are normalised to the
      overall number of generated events.} }
    \label{f:impos}
  \end{center}
\end{figure}
In order to determine the frequency that this might occur,
we generated for Figure~\ref{f:impos} \eeqq\ Monte Carlo events with
full hadronization, at $\sqrt{s}=91.2$ GeV. The full points present the fraction of events
that have multiple regions in \ycut\ with the same jet multiplicity,
as a function of the jet-multiplicity. 
For our generated events,
for example, about 2.2\% have multiple regions in \ycut\ that lead to a
four-jet final state, albeit with different jet four-momenta.  
In the same figure the open points show the
fraction of events were the indicated number of jets could not be
resolved. For example, in about 2.5\% of events no four-jet
configuration could be found.

The figure shows that the fraction of impossible jet multiplicities
and multiple jet multiplicities increases with increasing $n$, reaches
a maximum at around $n=12$, and decreases again. The generated events
have a mean total multiplicity of 44.2, and 95\% of the events have a
multiplicity larger than 25.  Note that both distributions are
naturally limited by the input number of four-momenta in each event.

%
\begin{figure}[!tb]
  \begin{center}
    \resizebox{\textwidth}{!}  {\includegraphics{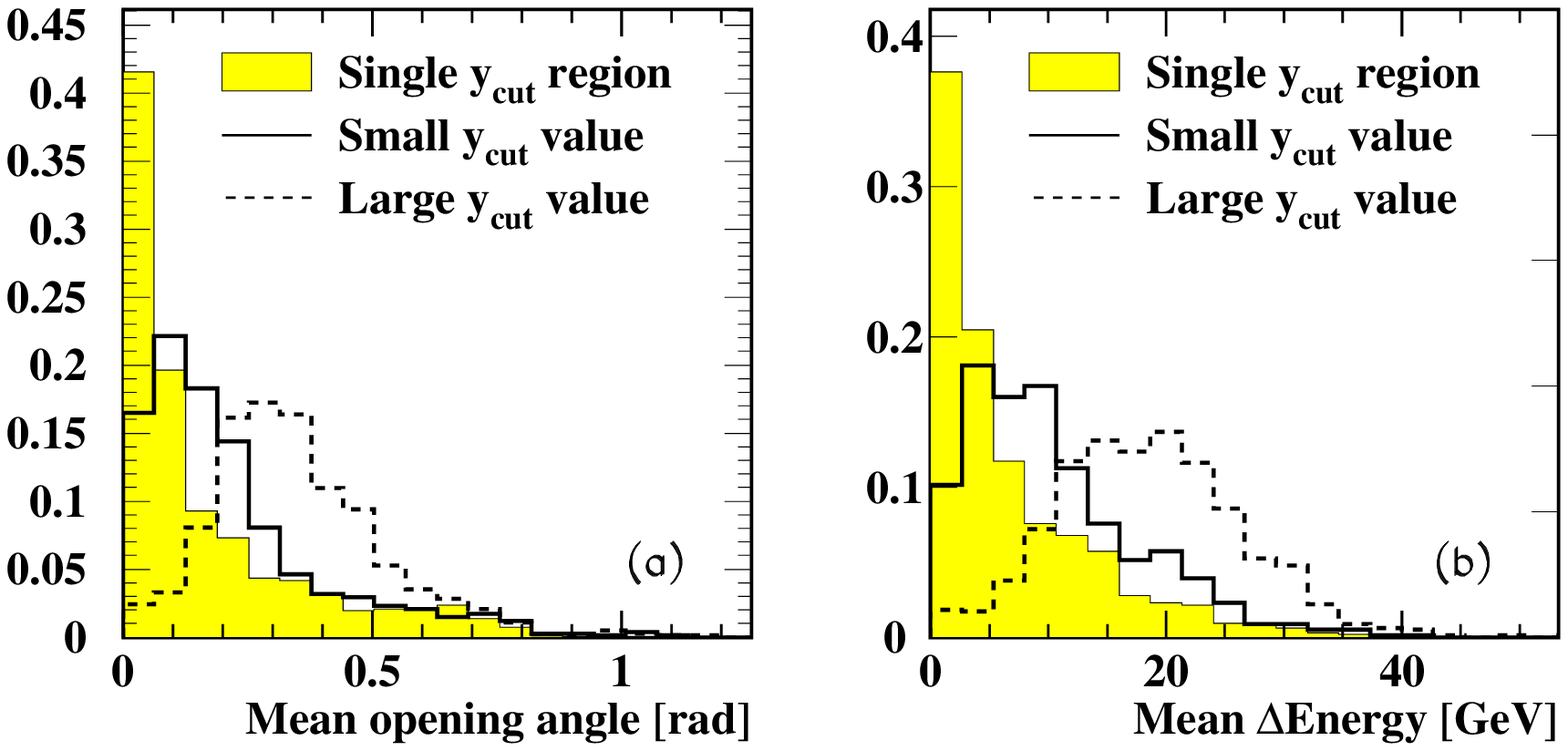}}
    \caption[]
    { {\sl Monte Carlo generated $W^+W^-$ events at $\sqrt{s} = 184$
        GeV.  In a) we present the mean of the opening angles between
        the momenta of the four partons and of the four jets at the
        hadron level.  The shaded histogram shows the events for which
        one four-jet configuration was found. The open histograms
        correspond to the set of events were the Cambridge algorithm
        returned two four-jet configurations.  The one with the dashed
        line corresponds to the configuration with the higher values
        of \ycut, whereas the one with the full line corresponds to
        the configuration with the lower values of \ycut. In b) the
        same is shown, but for the mean absolute energy difference
        between the four parton and hadron jets.  } }
    \label{f:fourj}
  \end{center}
\end{figure}
As another example we generated hadronic decays of $W^+W^-$ pairs
at LEP2~\cite{b:lep2ww} using PYTHIA: $\epem \rightarrow W^+W^- \rightarrow
q\bar{q}'q''\bar{q}'''$, at $\sqrt{s}=184$ GeV. Information about the
kinematics of the two $W$'s can be obtained by forcing the hadronic
final state to four jets. 

Using the Cambridge finder, we find that about 0.9\% of the events
have multiple regions in \ycut\ with four final state jets. For those
events one therefore has the freedom to select the set of jets with
the larger \ycut\ values, or the set with the smaller \ycut\
values. Clearly the selection which corresponds closer to the four
primary partons is preferred.  In Figure~\ref{f:fourj}a we compare the
mean opening angle between the jets and primary four partons, and in
Figure~\ref{f:fourj}b the mean absolute energy difference between the
jets and the primary partons.  It can be clearly seen that in both cases 
the resolution is better for the fraction of 99\% of events in which
only one four-jet configuration is found.  For the small fraction of
events where two four-jet configurations were found, the jet
configuration with the lower values of \ycut\ matches the four primary
partons better than the one with larger values of \ycut, which can be
explained by the following observation.  In the majority of events for
which the Cambridge algorithm returned two four jet configurations the
appearance of hard gluon radiation in the parton shower was observed.
Detailed inspection revealed that the hard gluon, radiated from a
quark-pair originating from one $W$, points in the direction of a
quark originating from the other $W$.  The configuration with the low
value of \ycut\ correctly separates the gluon from this quark by the
mechanism of {\em soft-freezing}, whereas they are merged for the
configuration with the larger value of \ycut. The correspondence
between partons and jets is therefore better in the configuration with
the lower value of \ycut.


\section{ Conclusions }
In this note we review the Cambridge jet clustering algorithm, as was
recently introduced in~\cite{b:cambridge}. We show some of its
particularities for Monte Carlo generated events. Firstly, the
algorithm may find several regions in \ycut\ with identical final
state multiplicity, but different jet four-momenta. Secondly, for some
events it is impossible to resolve a certain jet multiplicity.  Both
these properties are absent in the JADE and Durham algorithms.

We propose a fast, new algorithm that is able to determine the transition
values for \ycut, based on the {\tt YCLUS} package. All transition
values, jet multiplicities, jet four-momenta and the jet to particle
associations are derived and stored, and can be subsequently 
inferred for all values of \ycut\ without any substantial additional
computing time.

Using this algorithm we determine the hadronization corrections of
$\epem \rightarrow q\bar{q}$ generated events according to PYTHIA, by
comparing parton and hadron level values for \ytt, both for the Durham
and Cambridge algorithms. This comparative study of the two algorithms
is completed by a presentation of the relative jet production rates.
For a large interval of \ycut\ values the hadronization corrections
for the Cambridge algorithm are found to be significantly larger than
for the Durham algorithm. However, in the region of very small values
of \ycut\ ($\ycut < 10^{-3.2}$), the hadronization corrections are
large, but better under control for the Cambridge algorithm. Note that
for very low values of \ycut\ the reliability of the comparative Monte
Carlo studies is limited due to the fact that for these values of
\ycut\ the approximations used for the implementation of QCD in PYTHIA
might not give an appropriate description of the hadronization
process.


Further, we present for the Cambridge algorithm the fraction of events
for which certain jet multiplicity could never be resolved, or could
be resolved multiple times.  Four jet final states were explicitly
studied in hadronic decays of $W^+W^-$ events. The large fraction of
events where just one four jet configuration was found has better
energy and angular resolution than the small fraction of events with
multiple four jet configurations.


Fortran code, containing our {\tt CKERN} routines to obtain the
\ycut\ transition values, can be obtained from the World-Wide Web at

{\tt http://wwwcn1.cern.ch/\~{}stanb/ckern/ckern.html}.

\section{ Acknowledgements }
We would like to thank S.\ Bethke for help and inspiring discussions,
as well as B.\ Webber, Yu.\ Dokshitzer, S.\ Moretti and Z.\ Trocsanyi
for comments.  We like to thank CERN for its hospitality and in
particular the OPAL collaboration for providing indispensable
recourses.

\end{document}